\documentclass[twocolumn,english,aps,prl,floats,showpacs]{revtex4}
\usepackage[T1]{fontenc}
\usepackage[latin9]{inputenc}
\usepackage{babel}
\usepackage{amsmath}
\usepackage{amssymb}
\usepackage{graphicx}
\usepackage{esint}
\usepackage[unicode=true,
 bookmarks=true,bookmarksnumbered=false,bookmarksopen=false,
 breaklinks=false,pdfborder={0 0 1},backref=false,colorlinks=false]
 {hyperref}
\hypersetup{
 pdfauthor={D. A. Garanin},
 pdfkeywords={magnetic particle reversal microwave}}

\makeatletter

\newcommand{\lyxdot}{.}

\@ifundefined{textcolor}{}
{%
 \definecolor{BLACK}{gray}{0}
 \definecolor{WHITE}{gray}{1}
 \definecolor{RED}{rgb}{1,0,0}
 \definecolor{GREEN}{rgb}{0,1,0}
 \definecolor{BLUE}{rgb}{0,0,1}
 \definecolor{CYAN}{cmyk}{1,0,0,0}
 \definecolor{MAGENTA}{cmyk}{0,1,0,0}
 \definecolor{YELLOW}{cmyk}{0,0,1,0}
}

\usepackage{babel}

\makeatother

\begin{document}
\bibliographystyle{prsty}

\title{Turbulent fronts of quantum detonation in molecular magnets}

\author{D. A. Garanin}

\affiliation{Physics Department, Lehman College, City University of New York \\
 250 Bedford Park Boulevard West, Bronx, New York 10468-1589, USA}

\date{\today}
\begin{abstract}
Dipolar-controlled quantum deflagration going over into quantum detonation
in the elongated Mn$_{12}$ Ac molecular magnet in a strong transverse
field has been considered within the full 3$d$ model. It is shown
that within the dipolar window around tunneling resonances the deflagration
front is non-flat. With increasing bias, dipolar instability makes
the front turbulent, while its speed reaches sonic values, that is
a signature of detonation.
\end{abstract}

\pacs{75.50.Xx, 75.45.+j, 76.20.+q}

\maketitle
Magnetic burning or deflagration, similar to the usual chemical burning,
has been experimentally observed on elongated crystals of molecular
magnet (MM) Mn$_{12}$ Ac in Refs. \cite{suzetal05prl,heretal05prl}.
Deflagration is controlled by the temperature rising as the result
of the relaxation from the metastable to stable state of this bistable
system (burning), accompanied by the heat conduction toward the cold
unburned region. This leads to the formation of a flat burning front
moving at a constant speed controlled by the energy release set by
the bias magnetic field $B_{z}$ applied along the easy axis of magnetic
molecules. This discovery triggered theoretical \cite{garchu07prb}
and further experimental \cite{mchughetal07prb,hughetal09prb-tuning}
research. In these experiments no transverse field was applied and
the front speed was as low as 1-15 m/s. Fast moving burning fronts
in Mn$_{12}$ Ac initiated by a fast magnetic field sweep have been
observed in Ref. \cite{decvanmostejhermac09prl}. In this region,
deflagration can go over into detonation with a speed comparable to
the speed of sound \cite{modbycmar11prl}. Crossover from relaxation
to deflagration in Mn$_{12}$ Ac in a strong transverse field has
been experimentally studied in Ref. \cite{subedietal13prl}.

Since MM are famous exponents of the resonant spin tunneling \cite{frisartejzio96prl,heretal96epl}
(see Ref. \cite{gatsesvil06book} for a review), manifestations of
the latter were expected in magnetic deflagration. Indeed, maxima
of the front speed at tunneling resonances were seen in Refs. \cite{heretal05prl,mchughetal07prb}.
The simplest way to explain these maxima is to use, within the standard
deflagration theory, the Arrhenius relaxation rate $\Gamma(B_{z},T)$
having maxima at $B_{z}$ corresponding to tunneling resonances \cite{heretal05prl,garchu07prb}.

It was further suggested that in the case of a strong tunneling directly
out of the metastable ground state, caused by a strong transverse
magnetic field, fronts of non-thermal quantum or ``cold'' deflagration
are possible \cite{garchu09prl,gar09prb}. Here, instead of the temperature,
tunneling is controlled by the dipolar field of the crystal that can
block and unblock tunneling by setting magnetic molecules on or off
resonance. Within the simplified $1d$ approximation it was shown
that the magnetization in the moving front is adjusting self-consistently,
so that the dipolar field unblocks tunneling within the front core.
The theory of quantum deflagration has been generalized \cite{garjaa10prbrc,garsho12prb}
to include both quantum and thermal effects via the relaxation rate
$\Gamma(B_{z},T)$ numerically calculated from the density-matrix
equation \cite{gar12acp}. The main finding was a strong quantum acceleration
of the front within the dipolar window around the resonance values
of the external field $B_{z}=B_{k}$:
\begin{equation}
B_{k}-B_{z}^{(k_{D})}\leq B_{z}\leq B_{k}+B_{z}^{(D)},\label{eq:Dipolar-window}
\end{equation}
 where $B_{z}^{(D)}=52.6$ mT is the dipolar field produced by the
fully magnetized long crystal and $B_{z}^{(k_{D})}=72.9$ mT \cite{garsho12prb}.The
front speed was supersonic and apparently diverging towards the right
end of the dipolar window, see Fig. 15 of Ref. \cite{garsho12prb}
and discussion therein. Indeed, as the dipole-dipole interaction (DDI)
is instantaneous, there is no limitation on the front speed.

\begin{figure}
\centering\includegraphics[width=8cm]{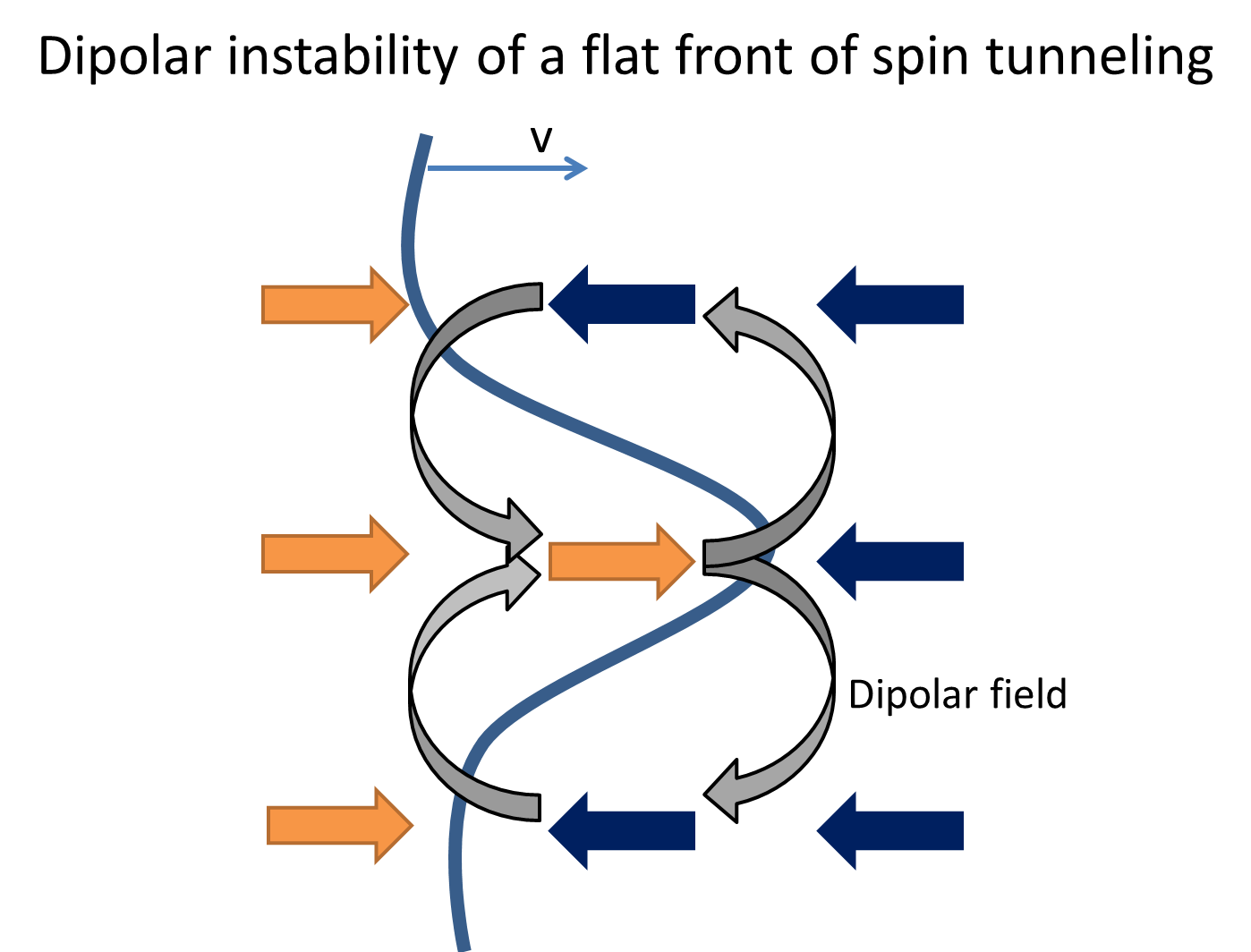}

\caption{(Color online). Dipolar instability of a smooth front of tunneling.
The leading parts of the front (center) are helped by the lagging
neighbors while hampering the motion of the latter.}

\label{Fig-dipolar_instability}
\end{figure}

While fast quantum deflagration in MM is avaiting experimental confirmation,
a theoretical problem remains. The $1d$ approximation assumes that
the magnetization and dipolar field in the front depend only on the
coordinate $z$ along the sample (that coincides with the MM easy
axis). This is a crude approximation, and there should be a dependence
on $x$ and $y$ that renders the front non-flat. Moreover, dipolar-controlled
front should be unstable at a small scale because portions of the
front that are ahead of the neighbors create dipolar field on the
latter that hampers their motion, while the lagging neighbors create
a field that accelerates the leader (see Fig. \ref{Fig-dipolar_instability}).
This instability reflects the fact that a long MM crystal tends to
split up into domains parallel rather than perpendicular to the $z$
axis \cite{gar10prbrc}. As the result, the front can become turbulent
with a chaotic dipolar bias rendering many magnetic molecules inside
the front core off resonance. The question is whether the front speed
drops to that of the regular slow burning \cite{suzetal05prl,heretal05epl}
or remains fast. In the latter case one has to speak of \textit{quantum
detonation} rather than of quantum deflagration.

The purpose of this Letter is to present a full $3d$ theory of fronts
of tunneling in molecular magnets and clarify the role of the dipolar
instability. A $3d$ model with DDI presents a numerical challenge,
still its solution is possible on modern workstations.

The system of equation describing deflagration with quantum effects
in molecular magnets \cite{garjaa10prbrc} includes the rate equation
describing relaxation of the metastable population $n$ ($1\leq n\leq1$)
\begin{equation}
\dot{n}(t,\mathbf{r})=-\Gamma\left[B_{\mathrm{tot},z}(\mathbf{r}),T(\mathbf{r})\right]\left[n(t,\mathbf{r})-n^{(\mathrm{eq})}\right].\label{ndot}
\end{equation}
Here the relaxation rate $\Gamma$ depends on the total bias field
\begin{equation}
B_{\mathrm{tot},z}(\mathbf{r})=B_{z}+B_{z}^{(D)}(\mathbf{r})\equiv B_{z}+\left(Sg\mu_{B}/v_{0}\right)D_{zz}(\mathbf{r}),\label{eq:Btot_z}
\end{equation}
 (external plus dipolar field) and the temperature at a given point,
while $n^{(\mathrm{eq})}(T)$ is the equilibrium metastable population
(set to zero below). Here $v_{0}=a^{2}c$ is the unit-cell volume,
$a$ and $c$ are lattice spacings, and $D_{zz}(\mathbf{r})$ is the
reduced dipolar field. Another equation is the heat conduction equation
that can be conveniently written in terms of the thermal energy $\mathcal{E}$
per magnetic molecule 
\begin{equation}
\mathcal{\dot{E}}(t,\mathbf{r})=\nabla\cdot\kappa\nabla\mathcal{E}(t,\mathbf{r})-\Delta E\dot{n}(t,\mathbf{r}).\label{Edot}
\end{equation}
 Here $\Delta E\cong2Sg\mu_{B}B_{z}$ in the source term of this equation
is the released energy per molecule, $\kappa$ is thermal diffusivity
(a crude estimate $\kappa\simeq10^{-5}$ m$^{2}$/s \cite{suzetal05prl,hughetal09prb-tuning}).
$\mathcal{E}$ and $T$ are related via the measured \cite{gometal98prb}
heat capacity $C=d\mathcal{E}/dT$. 

\begin{figure}
\centering\includegraphics[width=8cm]{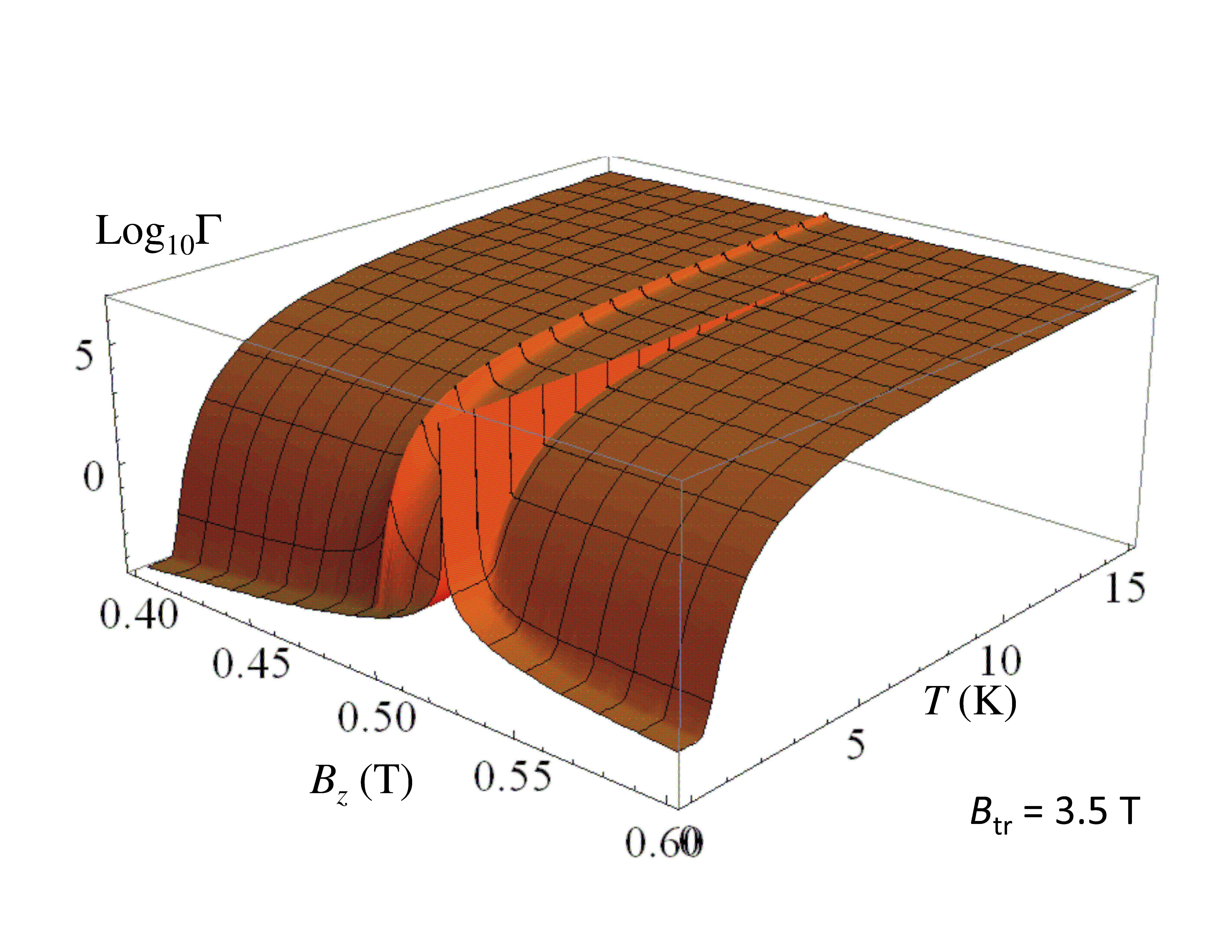}\caption{(Color online). Relaxation rate of Mn$_{12}$Ac vs temperature and
longitudinal magnetic field in the transverse field $B_{\bot}=3.5$
T. One can see the ground-state resonance at $B_{z}=0.522$T and the
first-excited-state resonance at $B_{z}=0.490$ T for $k=1$ multiplet.}

\label{Fig-3D-Gamma_Btr=00003D3.5T}
\end{figure}

While the Eqs. \ref{ndot} and \ref{Edot} are the same as the standard
deflagration equations \cite{garchu07prb}, the resonance form of
$\Gamma$ and its crucial dependence on the dipolar field $B_{z}^{(D)}(\mathbf{r})$
that is defined by $n(t,\mathbf{r})$ makes a big difference. For
the realistic model of Mn$_{12}$ Ac with the uniaxial part of the
Hamiltonian $-DS_{z}^{2}-AS_{z}^{4}$, tunneling resonances are achieved
at $B_{\mathrm{tot},z}=B_{km}$, where 
\begin{equation}
g\mu_{B}B_{km}=k\left[D+\left(m^{2}+(m+k)^{2}\right)A\right],\label{eq:Bzkm-Def}
\end{equation}
$k=0,1,2,\ldots$ and $m=-S,-S+1,\ldots$ label the metastable spin
states. The small fourth-order anisotropy $A$ splits $k$ resonances
into $m$ multiplets. With Mn$_{12}$ Ac parameters from Ref. \cite{barkenrumhencri03prl}
and using the density-matrix-equation method of Ref. \cite{gar12acp},
for the transverse field $B_{\bot}=3.5$ T one obtains $\Gamma$ around
the first resonance, $k=1,$ shown in Fig. \ref{Fig-3D-Gamma_Btr=00003D3.5T}.
At such strong $B_{\bot}$ only the resonances with $m=-10$ (ground-state
resonance) and $m=-9$ are seen, while all the other broadened away.

The most challenging part of the work is numerical calculation of
the dipolar field produced by the sample. $D_{zz}(\mathbf{r})$ in
Eq. (\ref{eq:Btot_z}) is defined by the lattice sum $D_{zz}(\mathbf{r}_{i})=\sum_{j}\phi(\mathbf{r}_{j}-\mathbf{r}_{i})\sigma_{z}(\mathbf{r}_{j})$,
where 
\begin{equation}
\phi(\mathbf{r})\equiv v_{0}\left[3\left(\mathbf{e}_{z}\cdot\mathbf{n}\right)^{2}-1\right]/r^{3},\qquad\mathbf{n}\equiv\mathbf{r}/r.\label{eq:phi-def}
\end{equation}
To calculate this sum, one can introduce a small sphere of radius
$r_{0}$ around $\mathbf{r}_{i}$ satisfying $v_{0}^{1/3}\ll r_{0}\ll L$,
where $L$ is the (macrocopic) size of the sample. Assuming that $\sigma_{z}(\mathbf{r})$
does not change at the scale $r_{0}$, one obtains 
\begin{equation}
D_{zz}(\mathbf{r})=\frac{\nu}{v_{0}}\int_{\left|\mathbf{r}'-\mathbf{r}\right|>r_{0}}d\mathbf{r}'\phi(\mathbf{r}'-\mathbf{r})\sigma_{z}(\mathbf{r}')+\sigma_{z}(\mathbf{r})\bar{D}_{zz}^{(\mathrm{sph})},\label{eq:Dzz-0}
\end{equation}
where $\nu=2$ is the number of magnetic molecules per unit cell $v_{0}$
for the body-centered tetragonal Mn$_{12}$ Ac and the lattice-dependent
$\bar{D}_{zz}^{(\mathrm{sph})}=2.155$ comes from summation over the
small sphere. The restriction on the integration can be removed by
subtracting $\sigma_{z}(\mathbf{r})$ from the integrand and adding
the counter-term. Finally one obtains
\begin{eqnarray}
D_{zz}(\mathbf{r}) & = & \frac{\nu}{v_{0}}\int d\mathbf{r}'\phi(\mathbf{r}'-\mathbf{r})\left(\sigma_{z}(\mathbf{r}')-\sigma_{z}(\mathbf{r})\right)\nonumber \\
 & + & \sigma_{z}(\mathbf{r})\left(\nu\bar{\mathcal{D}}_{zz}(\mathbf{r})-k_{D}\right),\label{eq:Dzz-3d-Def}
\end{eqnarray}
where $k_{D}\equiv8\pi\nu/3-\bar{D}_{zz}^{(\mathrm{sph})}=14.6$.
$\bar{\mathcal{D}}_{zz}(\mathbf{r})$ is the reduced magnetostatic
field given by an integral over the sample surface. For a rectangular
sample with dimensions $2L_{x}\times2L_{y}\times2L_{z}$ and $-L_{x}\leq x\leq L_{x}$
etc. one has \cite{garsch05prb} 
\begin{eqnarray}
 &  & \bar{\mathcal{D}}_{zz}(\mathbf{r})=\sum_{\eta_{x},\eta_{y},\eta_{z}=\pm1}\arctan\nonumber \\
 &  & \frac{\left(L_{x}+\eta_{x}x\right)^{-1}\left(L_{y}+\eta_{y}y\right)\left(L_{z}+\eta_{z}z\right)}{\sqrt{\left(L_{x}+\eta_{x}x\right)^{2}+\left(L_{y}+\eta_{y}y\right)^{2}+\left(L_{z}+\eta_{z}z\right)^{2}}}\label{eq:Dzz-3d-Surface}
\end{eqnarray}
$+\left(x\Rightarrow y\right)$, in total 16 different $\arctan$
terms. Inside a uniformly magnetized long Mn$_{12}$ Ac crystal Eq.
(\ref{eq:Dzz-3d-Def}) yields $D_{zz}=10.53$ that in real units corresponds
to $B_{z}^{(D)}=52.6$ mT. The same value of the dipolar field follows
from the experiment \cite{mchughetal09prb}.

For numerical solution, box-shape samples with dimensions $L_{x}\times L_{y}\times L_{z}$
and $0\leq x\leq L_{x}$ etc. are considered, and the problem is discretized
with a rectangular grid. This makes the integral in Eq. (\ref{eq:Dzz-3d-Def})
a sum that can be calculated using the fast Fourier transform (FFT).
In this work, numerical solution was implemented in Wolfram Mathematica,
whereas the dipolar sum was computed using FFT based and very fast
ListConvolve command. After discretization of the Laplace operator
in Eq. (\ref{Edot}) the system of equations for quantum deflagration
becomes a system of nonlinear ordinary differential equations. It
has been solved by the fixed-step compiled 4th-order Runge-Kutta method
and, preferred, by somewhat more efficient 5th-order Butcher's Runge-Kutta
method making 6 evaluations per step. Unfortunately, long-range DDI
prevents parallelization of the code. Instead of $\mathcal{E}$, it
is more convenient to use $\mathcal{E}/\Delta E+n$ as one of unknown
functions. The grid sizes were $64\times64\times300$, $100\times100\times200,$
and $100\times100\times300$. To reduce the amount of calculations,
symmetry conditions within $xy$ planes have been imposed. 

\begin{figure}
\centering\includegraphics[width=8cm]{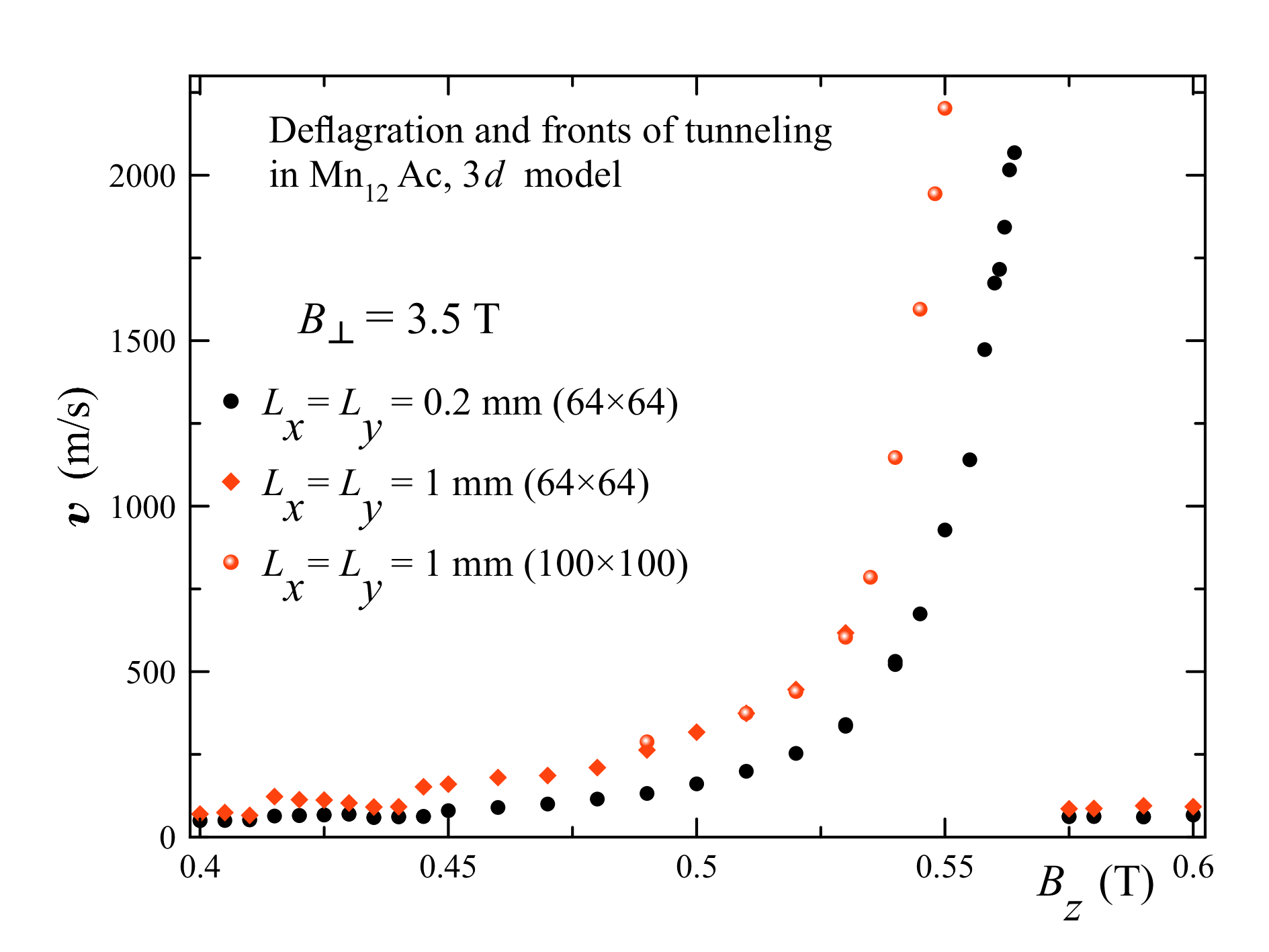}\caption{(Color online). Front speed within the $3d$ model for a strong transverse
field ($B_{\bot}=3.5$ T) in the vicinity of the ground-state tunneling
resonance at $B_{z}=B_{k}=0.522$ T.}

\label{Fig-v_Btr=00003D3.5T}
\end{figure}

Deflagration was ignited in the uniformly magnetized state $n=1$
at the left end of the crystal, $z=0$, by a quick temperature rise
at the boundary. At higher $B_{z}$, deeper into the dipolar window,
the resonance condition is fulfilled at a surface close to the left
end, so that quantum deflagration begins spontaneously there. The
right end of the sample was protected from spontaneous ignition by
adding a semi-infinite fictituous sample with a fixed ``down'' magnetization,
$n=1$. The sample was considered as thermally insulated, the initial
temperature being almost zero. The front speed was determined from
the time needed for the front to arrive at the right end or from the
time dependence of the average metastable population. Computations
were done for $L_{x}=L_{y}=0.2$ mm \cite{mchughetal07prb} and 1
mm \cite{heretal05prl}. One burning event required about one day
computations on a workstation.

The results for the front speed in the vicinity of the first tunneling
resonance ($k=1$) are shown in Fig. \ref{Fig-v_Btr=00003D3.5T}.
Beyond the dipolar window, $B_{z}<0.45$ T and $B_{z}>0.573$ T, there
is no tunneling and the usual temperature-driven deflagration with
a smooth flat front and $v\approx50$ m/s takes place. Entering the
dipolar window from the left leads to a gradual increase of the effect
of tunneling. The front speed increases and reaches sonic values,
then the front becomes so fast that numerical calculations require
a too long sample and become too difficult. Theoretically, the front
speed diverges at the right end of the dipolar window $B_{k}+B_{z}^{(D)}=0.522+0.0526=0.573$
T, where $B_{k}$ is the ground-state tunneling resonance and $k=1$.
Although the front speed in the $3d$ model is lower than in the $1d$
model, qualitatively the results are similar (cf. Fig. 15 of Ref.
\cite{garsho12prb}). Here one cannot see any contribution of the
excited-state tunneling resonance (peak at $B_{z}=0.490$ T in Fig.
\ref{Fig-3D-Gamma_Btr=00003D3.5T}), the whole tunneling effect being
due the the ground-state tunneling.

\begin{figure}
\centering\includegraphics[width=8cm]{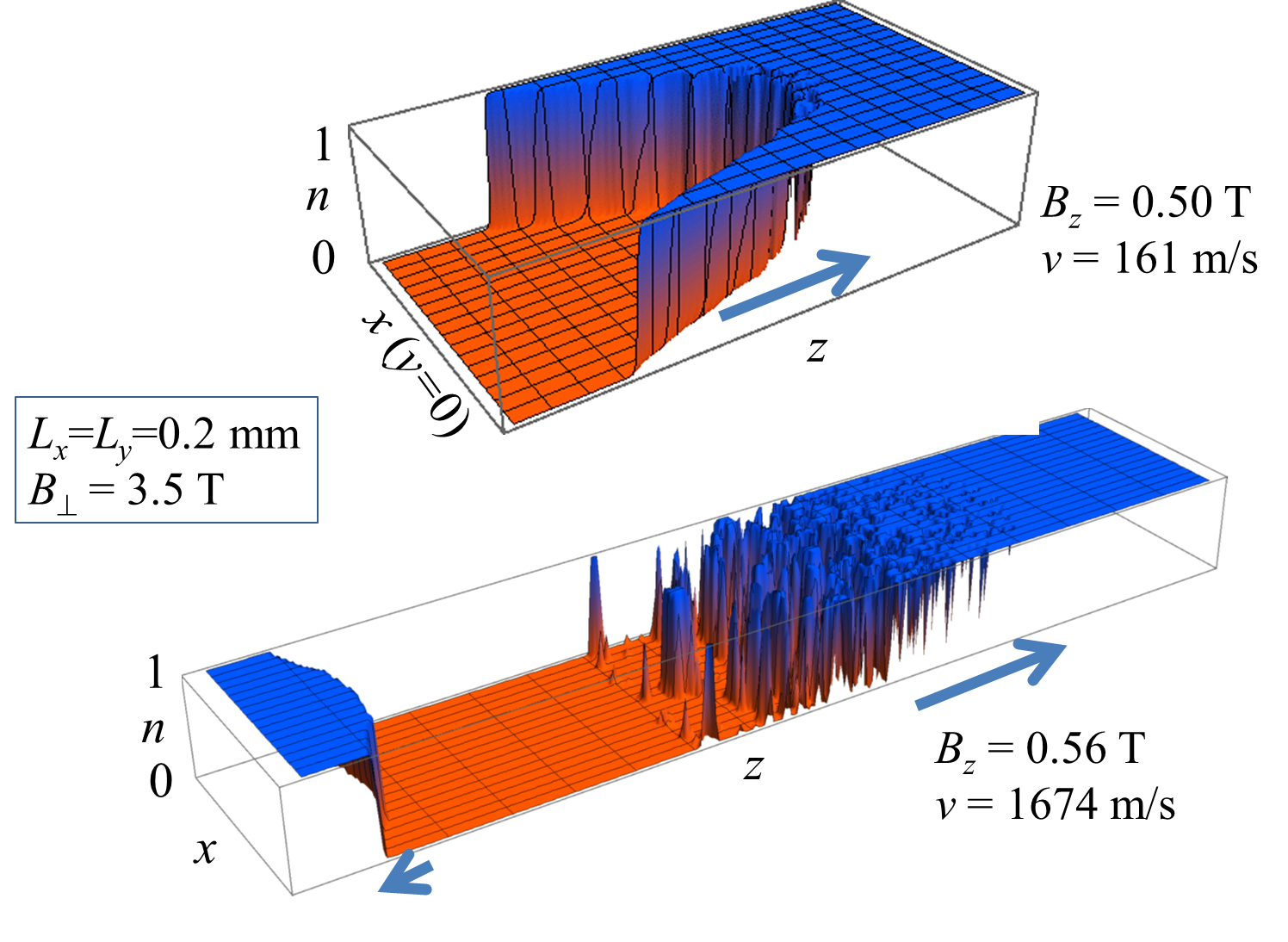}

\caption{(Color online). Profile of the metastable population $n$ in the $3d$
model of fronts of tunneling in a 0.2 mm Mn$_{12}$Ac crystal at $B_{\bot}=3.5$
T and $B_{z}=0.5$ T (upper) and 0.56 T (lower).}

\label{Fig-3d-front_profile_Lx=00003DLy=00003D0.2mm_Bz=00003D0.5_and_0.56T} 
\end{figure}

\begin{figure}
\centering\includegraphics[width=8cm]{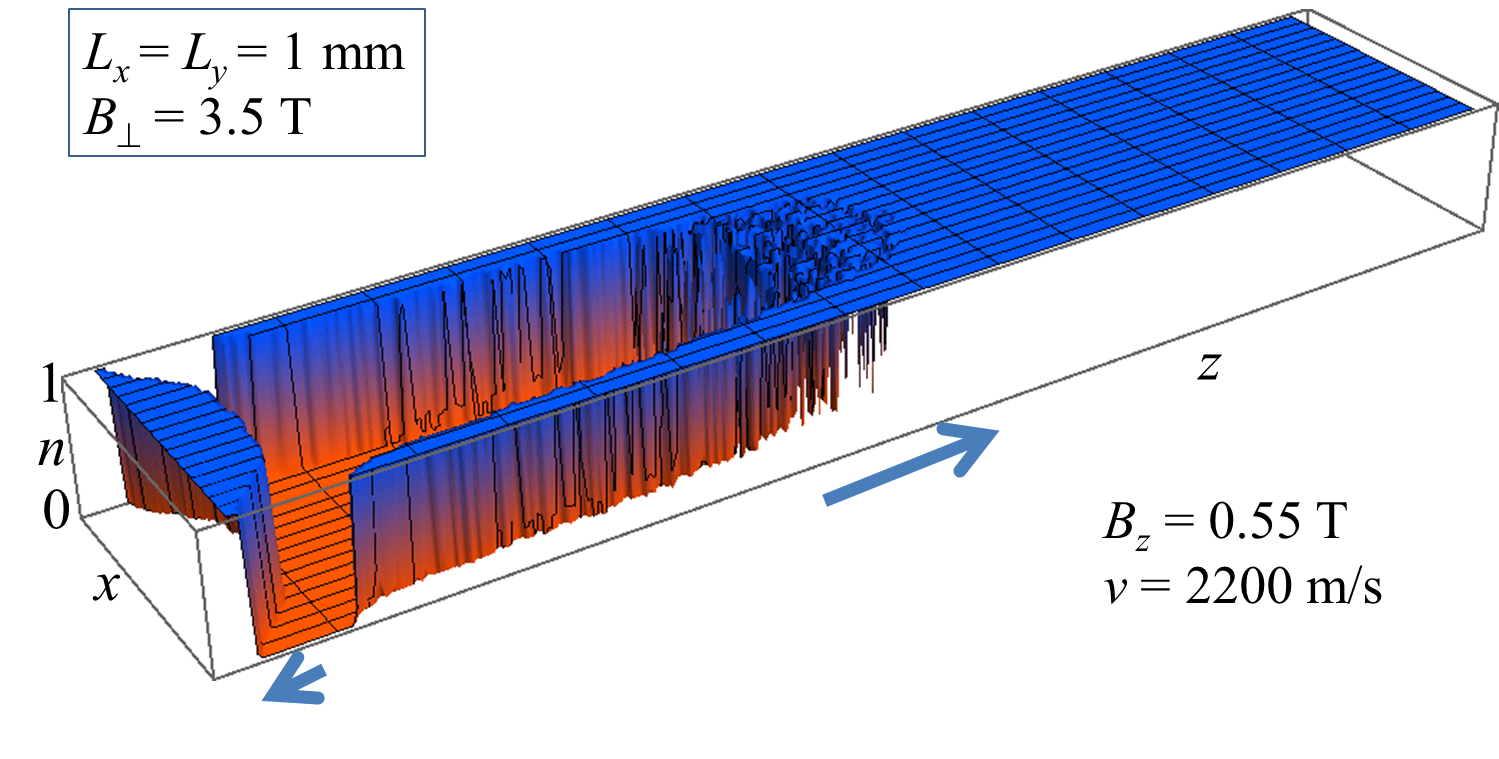}\caption{(Color online). Profile of the metastable population $n$ in the $3d$
model of quantum deflagration in a 1 mm Mn$_{12}$Ac crystal at $B_{\bot}=3.5$
T and $B_{z}=0.55$ T.}

\label{Fig-3d-front_profile_Btr=00003D3.5T_Bz=00003D0.55T_Ltr=00003D1mm_Lztil=00003D15000_64x64x300}
\end{figure}

The structure of the front of tunneling is shown in Fig. \ref{Fig-3d-front_profile_Lx=00003DLy=00003D0.2mm_Bz=00003D0.5_and_0.56T}.
Here $n$ is represented as a function of $x$ and $z$ at $y=0$.
Color coding is blue for the cold unburned regions ($n\approx1)$
and red for hot burned regions ($n\approx0$). Dipolar effect makes
the front progressively non-flat, its central part leading as the
speed is increasing, as seen in the upper image at $B_{z}=0.50$ T.
Also with increasing $B_{z}$ the front becomes gradually non-smooth.
However, thermal burning tends to smoothen the roughness created by
the dipolar instability discussed above as large temperature gradients
lead to increased heat conduction and equilibration along the front.
Whereas in the absence of thermal effect (cold deflagration) there
is a spatially irregular unburned metastable population behind the
front \cite{garchu09prl,gar09prb}, here a complete burning is achieved. 

Closer to the right border of the dipolar window the front becomes
turbulent and looking like precipitation, as seen in the lower image
in Fig. \ref{Fig-3d-front_profile_Lx=00003DLy=00003D0.2mm_Bz=00003D0.5_and_0.56T}
for $B_{z}=0.56$ T. Given its sonic speed that cannot be provided
by any thermal mechanism, one can speak of \textit{quantum detonation}.
The relaxation event in the figure began spontaneously on a surface
near the left end where the resonance condition was satisfied; From
there a regular slow-burning front is moving to the left and a quantum
detonation front is going to the right.

In thicker crystals, such as $L_{x}=L_{y}=1$ mm in Fig. \ref{Fig-3d-front_profile_Btr=00003D3.5T_Bz=00003D0.55T_Ltr=00003D1mm_Lztil=00003D15000_64x64x300},
the front speed is higher and non-flatness of the front is stronger
pronounced.

Although in the quantum detonation front the dipolar bias is wildly
changing in space so that the system does not stick to the resonance
(that was the main argument in the $1d$ theory of fronts of tunneling
\cite{garchu09prl,gar09prb}), tunneling is still very strong to ensure
sonic front speeds. This requires a closer investigation. Still, one
can argue that dynamic crossing the resonance can facilitate transitions
via Landau-Zener effect. 

To summarize, a full three-dimensional theory of dipolar-controlled
quantum deflagration with the temperature effect, going over into
quantum detonation, has been proposed. The quantum front can reach
sonic speeds near biased ground-state tunneling resonances and is
non-flat and turbulent. To activate quantum transitions out of the
metastable ground state, a strong transverse field has to be applied.
An experimental challenge is to prepare the initial state that has
a limited life time. 

This work has been supported by the NSF under Grant No. DMR-1161571.
The author thanks E. M. Chudnovsky for valuable discussions and help
in tuning the presentation.\bibliographystyle{apsrev}
\bibliography{gar-own,gar-relaxation,gar-tunneling,chu-own,gar-books}

\end{document}